\documentclass[preprint,12pt,authoryear,review]{elsarticle}

\usepackage{amssymb}

\usepackage{lineno}

\journal{Icarus}

\begin{document}

\begin{frontmatter}



\title{A common origin for dynamically associated near-Earth asteroid pairs}


\author[lowell]{Nicholas A. Moskovitz\corref{cor1}}
\cortext[cor1]{Corresponding author}
\ead{nmosko@lowell.edu} 

\author[czech,charles]{Petr Fatka}
\author[jpl]{Davide Farnocchia}
\author[lowell]{Maxime Devog\`{e}le}
\author[weiz]{David Polishook}
\author[nau]{Cristina A. Thomas}
\author[lowell]{Michael Mommert}
\author[ua]{Louis D. Avner}
\author[mit]{Richard P. Binzel}
\author[lowell]{Brian Burt}
\author[ua]{Eric Christensen}
\author[mit]{Francesca DeMeo}
\author[ucf]{Mary Hinkle}
\author[cfa]{Joseph L. Hora}
\author[nau]{Mitchell Magnusson}
\author[leidos]{Robert Matson}
\author[mit]{Michael Person}
\author[lowell]{Brian Skiff}
\author[lowell]{Audrey Thirouin}
\author[nau]{David Trilling}
\author[lowell]{Lawrence H. Wasserman}
\author[uh]{Mark Willman}

\address[lowell]{Lowell Observatory, 1400 W Mars Hill Road, Flagstaff, AZ 86001}
 
\address[czech]{Astronomical Institute, Academy of Sciences of the Czech Republic, Fri\v{c}ova 1,CZ-25165 Ond\v{r}ejov, Czech Republic}
\address[charles]{Institute of Astronomy, Charles University, Prague, V Hole\v{s}ovi\v{c}k\'{a}ch 2,CZ-18000 Prague 8, Czech Republic}
\address[jpl]{Jet Propulsion Laboratory, California Institute of Technology, 4800 Oak Grove Drive, Pasadena, CA 91109}
\address[weiz]{Faculty of Physics, Weizmann Institute of Science, 234 Herzl St. Rehovot 7610001, Israel}
\address[nau]{Department of Physics and Astronomy, Northern Arizona University, PO Box 6010, Flagstaff, AZ 86011}
\address[ua]{University of Arizona, Lunar and Planetary Laboratory, 1629 E. University Blvd., Tucson, AZ 85721}
\address[mit]{Department of Earth, Atmospheric \& Planetary Sciences, Massachusetts Institute of Technology}
\address[ucf]{University of Central Florida, Department of Physics, 4111 Libra Drive, Physical Sciences Bldg. 430, Orlando, FL 32816}
\address[cfa]{CfA Harvard-Smithsonian, 60 Garden St, Cambridge, MA 02138}
\address[leidos]{Leidos, 3030 Old Ranch Pkwy., Suite 200, Seal Beach, CA 90740}
\address[uh]{University of Hawaii, Institute for Astronomy, 34 `$\bar{O}$hi`a K$\bar{u}$ St., Pukalani, HI 96768 
USA }

\begin{abstract}
Though pairs of dynamically associated asteroids in the Main Belt have been identified and studied for over a decade, very few pair systems have been identified in the near-Earth asteroid population. We present data and analysis that supports the existence of two genetically related pairs in near-Earth space. The members of the individual systems, 2015 EE7 -- 2015 FP124 and 2017 SN16 -- 2018 RY7, are found to be of the same spectral taxonomic class, and both pairs are interpreted to have volatile-poor compositions. In conjunction with dynamical arguments, this suggests that these two systems formed via YORP spin-up and/or dissociation of a binary precursor. Backwards orbital integrations suggest a separation age of $<10$ kyr for the pair 2017 SN16 -- 2018 RY7, making these objects amongst the youngest multiple asteroid systems known to date. A unique separation age was not realized for 2015 EE7 -- 2015 FP124 due to large uncertainties associated with these objects' orbits. Determining the ages of such young pairs is of great value for testing models of space weathering and asteroid spin-state evolution. As the NEO catalog continues to grow with current and future discovery surveys, it is expected that more NEO pairs will be found, thus providing an ideal laboratory for studying time dependent evolutionary processes that are relevant to asteroids throughout the Solar System.

\end{abstract}

\begin{keyword}
Asteroids \sep Asteroids, dynamics \sep Near-Earth objects


\end{keyword}

\end{frontmatter}


%
%
\section{Introduction}
\label{intro}

The non-random distribution and clustering of asteroid orbital elements in the Main Belt has long been associated with the collisional disruption of parent body precursors \citep{hirayama18}. As the population of discovered asteroids continues to grow, finer scale clustering has been recognized including small, young families \citep[e.g.][]{nesvorny08,pravec18,carruba18} and pairs of asteroids on very similar heliocentric orbits \citep{vok08,pravec09,rozek11}. The proximity of these unbound asteroid pairs in orbital element space is inconsistent with a random distribution, and suggests a common origin for the larger primary and smaller secondary components. Backwards orbital integrations suggest that these pairs separated recently, in most cases less than $\sim$1 Myr ago \citep[e.g.][]{vok09,vok09b,pravec09,zizka16}. A shared origin for asteroid pairs is supported by the similar photometric colors and/or spectral properties of the two components \citep{moskovitz12,duddy13,polishook14,wolters14}. Several hundred candidate asteroid pair systems have been identified in the Main Belt \citep{pravec09}.

The search for clusters or pairs of minor planets amongst near-Earth objects (NEOs) has revealed fewer associations than in the Main Belt. Many of the dynamical associations identified in near-Earth space are between meteor streams and their parent comets \citep{jenniskens06}. Attempts to identify NEO families have been mostly unsuccessful \citep{fu05,schunova12}, likely due to the short coherence time of orbits in near-Earth space \citep{schunova14}. However, several exceptions have been found \citep{kasuga19}. Perhaps the best known is 3200 Phaethon and 155140 (2005 UD), which are believed to share a common origin based on the similarity of their orbital elements \citep{ohtsuka06}. The spectral similarity of these objects and the fact that they are both of the rare B-type spectral class further supports a connection \citep{jewitt06,kinoshita07}. A third object, 225416 (1999 YC), has also been linked to Phaethon \citep{ohtsuka08,kasuga08}. The formation of this dynamically associated system of asteroids is believed to be related to the Geminid meteor complex, which also shares an orbital similarity \citep{ohtsuka06}. Phaethon displays low levels of mass loss associated with perihelion passages. Phaethon's small perihelion distance (0.14 AU) and its consequently high surface temperatures are thought to cause thermal fracturing and/or cracking driven by volatile loss \citep{jewitt13}. It is unclear to what extent the thermal extremes experienced by Phaethon and its potentially volatile-rich composition (as evidenced by its B-type spectrum) contributed to the formation of its putative companions 2005 UD and 1999 YC. The mass ratios of Phaethon to 2005 UD and to 1999 YC are consistent with models for the formation of this system via rotational fission \citep{hanus18}.

A second pair of dynamically associated asteroids in near-Earth space are the objects 1566 Icarus and 2007 MK6 \citep{ohtsuka07}. These objects have been associated with the Taurid-Perseid meteor complex. Like Phaethon these objects also have low perihelia ($\sim0.19$ AU), but unlike Phaethon, Icarus is identified as an S-type asteroid \citep{chapman75}, suggesting a volatile-poor composition that might imply a different formation mechanism relative to objects in the Phaethon association.

A final example, based on recent dynamical analysis and very much related to the results presented here, is the pair 2017 SN16 and 2018 RY7 \citep{marcos19}. Beyond astrometry, no physical data have been reported on these objects. The age of this system and their mechanism of formation have not previously been investigated.

The formation of asteroid pairs almost certainly involves several different physical mechanisms that may not necessarily be relevant to every system or dynamical population. These mechanisms include collisional disruption of larger parent bodies, separation of gravitationally bound binary systems, YORP induced spin-up and subsequent rotational fissioning, thermally-driven break-up possibly associated with volatile loss, and tidal disruption of bodies on planet-crossing orbits \citep{richardson98,vok08,jacobson11,granvik16}. The combination of rotational properties and inferred relative mass ratio of pairs in the Main Belt are largely consistent with formation via rotational fission \citep{pravec10}. Similarity in the spin axis orientation of pair components also supports this claim \citep{polishook14b,vok17,pravec19}. However, the young pair 3749 Balam and 2009 BR60 in the Main Belt suggest that this process is indeed complicated -- Balam is also known to host an inner satellite on a nearly circular orbit as well as a more distant, more eccentric outer satellite \citep{marchis08,vok09b}. Interpreting the formation of any given pair system is further complicated by the fact that, due to issues of observational incompleteness, only the two largest members in a multi-body cluster of objects may be known.

Our aim here is to characterize and provide an interpretation for the origin of two newly recognized pairs in the NEO population. The asteroids 2015 EE7 (primary, hereafter EE7) and 2015 FP124 (secondary, hereafter FP124) were discovered within two weeks of one another in March of 2015. The asteroids 2017 SN16 (primary, hereafter SN16) and 2018 RY7 (secondary, hereafter RY7) were discovered roughly a year apart, but both had favorable apparitions in late 2018. All four objects were discovered by the Catalina Sky Survey \citep{christensen18}. The basic properties and orbital elements of these asteroids are given in Table \ref{tab.orb}. The NEOs in these systems have remarkably similar orbits, even down to a similarity in their mean anomaly, which accounts for their coincident windows of observability. The NEOLegacy survey on the Spitzer Space Telescope \citep{trilling16} observed EE7 and determined an albedo and diameter of 0.37 and 170 meters respectively (see \S\ref{sec.obs} for more details). \citet{warner15} obtained lightcurve photometry of EE7 across three different nights and suggested a rotation period of either 9.42 or 4.71 hours. Beyond this albedo and rotational information, no other physical characterization data have been published on the pair EE7 -- FP124. No physical characterization data exist for the pair SN16 -- RY7.

\begin{table}[h]
\caption{NEO pairs absolute magnitude (H), diameter (D), and orbital elements semi-major axis ($a$), eccentricity ($e$), inclination ($i$), longitude of ascending node ($\Omega$), argument of perihelion ($\omega$), and mean anomaly ($M$) at epoch 2458500.5 from Lowell Observatory's astorb database (asteroid.lowell.edu).
$^1$Diameter for EE7 from the Spitzer Space Telescope NEOLegacy survey \citep{trilling16}. Diameter for FP124 calculated assuming the same albedo (0.37) as EE7. Diameters for SN16 and RY7 calculated assuming the average albedo of 0.42 for V-type NEOs \citep{thomas11}.}
\begin{center}
\begin{tabular}{lllllllll}
\hline
Object	& H (mag)		& D (m)\footnote	& $a$ (AU)	& $e$	& $i$	& $\Omega$	& $\omega$	& $M$ \\
\hline
2015 EE7		& 20.2	& 170$^{+60}_{-30}$	& 1.702	& 0.411	& 27.31$^\circ$	& 9.41$^\circ$	& 219.2$^\circ$	& 238.9$^\circ$\\
2015 FP124	& 22.2	& $\sim80$		& 1.712	& 0.414	& 27.41$^\circ$	& 9.41$^\circ$	& 219.2$^\circ$	& 236.3$^\circ$\\
\hline
2017 SN16	& 23.2	& $\sim50$	& 1.016	& 0.146	& 13.38$^\circ$	& 2.74$^\circ$	& 137.0$^\circ$	& 341.7$^\circ$\\
2018 RY7		& 24.5	& $\sim25$		& 1.016	& 0.147	& 13.35$^\circ$	& 2.82$^\circ$	& 136.8$^\circ$	& 344.3$^\circ$\\
\hline
\end{tabular}
\end{center}
\label{tab.orb}
\end{table}%

Here we present new observations of these NEOs (\S\ref{sec.obs}) and use these data to place constraints on their physical properties (\S\ref{sec.properties}). Detailed dynamical analyses (\S\ref{sec.dyn}) provide insights into these pair's past orbital history, and in the case of SN16 -- RY7 we place a specific constraint on their separation age. The combination of physical properties and dynamical history provides insights into the formation mechanism of these objects (\S\ref{sec.discussion}). A summary at the end of this work enumerates broader implications for detection and characterization of genetically related pairs in near-Earth space (\S\ref{sec.summary}).

%
%
\section{Observations \& Data Reduction}
\label{sec.obs}

New photometric and spectroscopic observations of the two pair systems were obtained and interpreted along with archival lightcurve data of EE7 from Palmer Divide Station \citep{warner15}. Table \ref{tab.obs} provides an overview of the observational circumstances for each of these data sets.

\begin{table}[h]
\caption{Observational circumstances including apparent V-band magnitude ($V$), heliocentric distance ($r$), geocentric distance ($\Delta$), solar phase angle ($\alpha$), and observer centered ecliptic longitude ($\lambda$) and latitude ($\beta$) for NEO pairs 2015 EE7 -- 2015 FP124 and 2017 SN16 -- 2018 RY7. Palmer Divide observations of 2015 EE7 from \citet{warner15}.}
\begin{center}
{\tiny
\begin{tabular}{llllllllll}
\hline
		& 		& 					& 				& $V$	& $r$		& $\Delta$ 	& $\alpha$	& $\lambda$	& $\beta$ \\
Object	&UT Date& Telescope/Instrument	& Data Product		& [mag]	& [AU]	& [AU] 		& [deg]			& [deg]		& [deg] \\
\hline
2015 EE7	& 2015 Mar. 23	& Palmer Divide Station		& V-band photometry			& 18.0	& 1.120	& 0.161	& 37 		& 159	& 38\\
2015 EE7	& 2015 Mar. 24	& Palmer Divide Station		& V-band photometry			& 17.9	& 1.116	& 0.154	& 37 		& 158	& 36\\
2015 EE7	& 2015 Mar. 26	& Palmer Divide Station		& V-band photometry			& 17.7	& 1.107	& 0.140	& 36 		& 157	& 31\\
2015 EE7	& 2015 Mar. 31	& Lowell Perkins 1.8-m/PRISM	& R- and VR-band photometry	& 17.2	& 1.086	& 0.112	& 37 			& 151	& 15 \\
2015 EE7	& 2015 Apr. 01	& Lowell Perkins 1.8-m/PRISM	& VR-band photometry		& 17.1	& 1.082	& 0.109	& 38 			& 150	& 11 \\
2015 EE7	& 2015 Apr. 12	& Gemini-S/GMOS			& Visible spectra			& 17.9	& 1.044	& 0.115	& 65 			& 137	& -40 \\
2015 EE7	& 2015 Apr. 28	& SOAR/Goodman			& r-band photometry		& 19.9	& 1.009	& 0.227	& 83 				& 109	& -72 \\
2015 FP124	& 2015 Apr. 06	& SOAR/Goodman		& R-band photometry		& 19.2	& 1.044	& 0.089	& 59 			& 228	& -57 \\
2015 FP124	& 2015 Apr. 06	& SOAR/Goodman		& Visible spectra			& 19.2	& 1.045	& 0.089	& 58			& 228	& -57 \\
\hline
2017 SN16	& 2018 Oct. 02	& Gemini-S/GMOS		& Visible spectra			& 20.1	& 1.107	& 0.120	& 27			& 339	& 6 \\
2018 RY7		& 2018 Sep. 29	& Gemini-S/GMOS		& Visible spectra			& 20.3	& 1.108	& 0.107	& 7			& 1		& 6 \\
\hline
\end{tabular}
}
\end{center}
\label{tab.obs}
\end{table}%

\subsection{2015 EE7}
\label{sec.ee7}

Photometric observations of EE7 were obtained from Lowell Observatory's 1.8-m Perkins telescope with the Perkins Re-Imaging System (PRISM) on 31 March 2015 and 1 April 2015. PRISM has a 13.65' square field of view sampled at an image scale of 0.39" per pixel. The instrument was binned 3x3 for these observations. Images were obtained with a Cousins R filter and a broad VR filter, the later of which provides $>$50\% transmission between 525 and 700 nm, roughly spanning V- and R-bands. Exposure times across both nights ranged from 15 to 20 seconds.

Images of EE7 were also collected with the Goodman Spectrograph and Imager \citep{clemens04} at the Southern Astrophysical Research (SOAR) telescope on Cerro Pach\'on in central Chile. Goodman images a 7.2' circular field of view with a 4k x 4k Fairchild CCD at an image scale of 0.15 "/pixel. Images were binned 2x2. These data were taken on 28 April 2015 with an SDSS r' filter and span about 2.85 hours. Exposure times were 30-35 seconds.

Reduction of our images involved standard flat field and bias correction followed by measuring the asteroid photometry with the Photometry Pipeline developed by \citet{mommert17}. In short, this pipeline performed automatic image registration based on the Gaia DR1 catalog \citep{gaia16}, extraction of point source photometry using SourceExtractor \citep{bertin96}, photometric zero point calibration based on matching field stars to the PanSTARRS DR1 catalog \citep{flewelling16} for the Perkins data and to the SkyMapper DR1 catalog \citep{wolf18} for the SOAR data, and finally extracting the calibrated asteroid photometry by querying the JPL Horizons system \citep{giorgini97} for the position of the asteroid in each field. Through a curve of growth analysis the pipeline determined an optimal aperture of 4.6 binned pixels (5.4") in diameter for both nights of Perkins data and 3.7 binned pixels (1.1") in diameter for the SOAR data. All data were obtained in clear conditions with seeing stable at the $\sim25\%$ level.

The full collection of analyzed photometry for EE7 includes our R and VR images from Lowell's 1.8-m, our SDSS r' images from SOAR, and photometry from \citet{warner15} that was obtained unfiltered but calibrated to a zero point in Johnson V band. To derive a rotational lightcurve from these data, color corrections were made to bring all of the photometry (V, VR, SDSS r' filter data) to a common system (R). On 31 March 2015 we obtained about 10 minutes of VR images before switching to R-band. Minimizing the difference between linear interpolations of the data from these two filters resulted in a derived VR - R zero point offset for EE7 = 0.22. Based on our spectral data (\S\ref{sec.spec}) we know that EE7 is an Sq type asteroid in the \citet{bus02} system, thus a reasonable approximation for its color is V - R = 0.45 \citep{dandy03}. Lastly we use the r'-R transformation from \citet{jordi05} to translate our SOAR data. The color corrections were applied to the photometry before running the lightcurve analysis in \S\ref{sec.rot}.

Spectroscopic observations of EE7 were obtained from Gemini South with the Gemini Multi-Object Spectrograph (GMOS) on UT 12 April 2015. The instrument was configured with a 400 line per mm grating, a 455 nm blocking filter, and a 2" slit. The instrument has three 2k x 4k Hamamatsu chips arranged in a row to capture a wavelength range from about 0.45 - 0.95 $\mu m$ at an un-binned resolution of 0.7 \AA~per pixel. A total of 6 x 300 second exposures were obtained with the object dithered spatially along the slit at three different locations. In addition, the grating angle was adjusted midway through the exposures to offset the dispersion coverage and thus avoid loss of signal in the gaps between chips. Solar analog star SA 98-978 was similarly observed with 6 x 1-second exposures to enable telluric correction and to remove the solar continuum from the measured asteroid reflectance. The asteroid was observed at an airmass range of 1.01-1.08, and the solar analog from 1.25-1.26. The Gemini facility QTH (Quartz, Tungsten, Halogen) calibration lamp was used for flat field exposures, which were paired with the target spectra at each grating setting. CuAr arc lamp exposures from UT 14 April 2015 were used for determining the instrument dispersion solution. Reduction of these data were attempted in two different ways. First, we employed the Gemini IRAF package for GMOS spectral reduction. This package bias subtracts, cosmic ray cleans, flat fields, assigns a dispersion solution, extracts the 1D spectra, and combines the multiple exposures into a single spectrum.  Second, we employed a new custom-built python-based reduction package for single-order, long slit spectrographs like GMOS. This pipeline will ultimately be publicly released and the focus of a future paper. It was designed to effectively mimic the standard spectral reduction steps in IRAF. No significant differences were noted between these two reduction procedures, which provides a nice validation of this new tool. The final reduced spectrum of EE7, re-sampled to approximately 0.01 $\mu m$ per data point and normalized at 0.55 $\mu m$, is shown in Figure \ref{fig.spec_ee7}. Before computing the re-sampled value in each bin, a sigma-clipping algorithm was applied to remove points 2.5 sigma or more from the mean in each bin. The error bars on the final binned spectral points are equal to 1 standard deviation of the original data in each bin. 

EE7 also happened to be observed with the Spitzer Space Telescope in the framework of the NEOLegacy program, which aims at measuring Near-Earth asteroid diameters and albedos from their thermal emission \citep{trilling16}. From observations obtained centered on 2017-07-06 at 12:09 UT in Spitzer IRAC's Channel 2 \citep{fazio04} a diameter of $0.17_{-0.03}^{+0.06}$ km and a geometric albedo of $0.37_{-0.17}^{+0.20}$ were derived (data obtained from http://nearearthobjects.nau.edu/). These data consisted of $80 \times 100$ second exposures.

\subsection{2015 FP124}
\label{sec.fp124}

Observations of FP124 were obtained from the SOAR Telescope with the Goodman Spectrograph and Imager on 6 April 2015, approximately one week after the object was officially designated by the Minor Planet Center. These observations were already scheduled as part of the Mission Accessible Near-Earth Object Survey \citep[MANOS,][]{thirouin16,devogele19}. This timing was critical as FP124 was observable for physical characterization (V$<21$) for only about 2 weeks after its discovery. Both visible wavelength spectra and images in the Bessell R filter were obtained with this instrument. We employed 2x2 binning for these imaging observations. An exposure time of 7 seconds was used over the span of about two hours. Unfortunately, conditions were poor with heavy extinction and high sky background, resulting in only intermittent collection of useful images. Processing of the Goodman images followed the same procedure as that for EE7, with the exception that the Photometry Pipeline achieved photometric calibration using the SkyMapper DR1 catalog \citep{wolf18}.

The SOAR spectra of FP124 employed Goodman's 400 line per mm grating and a 3.2" slit. This grating produces a useful spectral range of approximately 0.5 - 0.9 $\mu m$ at a dispersion of about 1 \AA~per pixel. A total of 8 x 300 second exposures were obtained. Solar analog star SA 105-56 was observed for telluric and solar continuum correction. A total of 5 x 2-second exposures were obtained of the solar analog. The asteroid was observed at an airmass range of 1.32-1.4, and the solar analog at an airmass of 1.29. HgAr arc lamp exposures were obtained to determine a dispersion solution. No flat field correction was applied to the data. Attempts were made to use flat fields taken of an internal Quartz lamp, however unexplained non-uniformity in these flat fields introduced significant scatter into the final spectra. Our unsuccessful attempts to use these internal flats have ultimately led to a revision of the SOAR facility approach to flat-fielding Goodman data with the 400 line per mm grating. 

We do not expect the omission of flat field correction to have a large impact on the final spectrum. Other objects with well established spectral properties (e.g. NEO 1627 Ivar) also observed with Goodman without proper flat field correction turn out no different from archival data taken at other facilities. In later data sets from the same instrument when we did collect usable flat field exposures (using an external dome flat lamp) we find that the Goodman flat field is very uniform with no large scale gradients along the dispersion axis and only percent-level variability along the full extent of the spatial axis. Since the spectra of the asteroid and the solar analog were placed on very similar locations on the detector, and then were divided, we expect that first order flat field effects would cancel out. As such, we do not expect any large scale gradients to be imposed upon the data that would influence the final taxonomic classification, e.g. causing an S-type asteroid to appear as a Q-type or vice versa. Variations that we do see in the proper flat fields are at the sub-percent level and of relatively high spatial frequency relative to the broad gradients in which we are interested for determining spectral taxonomy. In this case the low signal-to-noise of the FP124 spectrum dominates the scatter in the final spectrum, such that any effects of the sub-percent flat field variability are lost in the noise.

Reduction of these data employed standard IRAF routines within the {\it apextract} package. However, the Goodman camera displayed significant fringing at wavelengths greater than 0.7 microns. To remove the fringing in the extracted spectrum, we performed a wavelet analysis using the python {\it pywt} package. We used a discrete wavelet transform within the Haar wavelet family, the signal processing was completed using a soft thresholding, and the universal threshold was applied to the coefficients \citep[see][for more details]{downie98}. We performed this wavelet analysis on the asteroid spectrum after we applied the solar analog for solar and telluric correction. We also tried the wavelet analysis on the asteroid and solar standard individually, but found no marked difference.

The final reduced spectrum of FP124, re-sampled to approximately 0.01 $\mu m$ per data point, is shown in Figure \ref{fig.spec_ee7}. Due to the lower signal-to-noise relative to the EE7 data, a more aggressive 2-sigma threshold was used in clipping the FP124 spectrum. Again error bars on the binned points represent one standard deviation of the original data in each bin.

\subsection{2017 SN16 \& 2018 RY7}
\label{sec.sn16}

We obtained visible spectra from Gemini South of SN16 and RY7. SN16 was observed on 2 October 2018, and RY7 was observed on 29 September 2018 about two weeks after its discovery. Both objects were observed with GMOS-S with exposure times and instrument settings identical to those used for 2015 EE7 (\S\ref{sec.ee7}). For RY7, only 3 of the 6 spectra obtained were found to be useful due to low signal-to-noise likely caused by object drift off of the slit and/or due to background field star contamination of the asteroid spectrum. Thus the resulting signal-to-noise of the combined RY7 spectrum is lower than desired, but as discussed in \S\ref{sec.spec}, is sufficient for our analysis purposes. Observations of the solar analog star SA 115-271 from each night were used to calibrate the asteroid spectra. SN16 was observed at an airmass range of 1.17-1.24 with the accompanying observations of the analog at 1.31-1.34. RY7 was observed at an airmass range of 1.24-1.33 with its analog at an airmass of 1.17. The final reduced spectra of SN16 and RY7 are shown in Figure \ref{fig.spec_sn16}, binned to approximately 0.01 $\mu m$ and 0.04 $\mu m$ per data point respectively and normalized at 0.55 $\mu m$. Sigma clipping with a 2.5 sigma threshold was applied to each spectrum. Error bars were computed in the same manner as before. 

To reduce uncertainty on the orbital elements of SN16 and RY7 we obtained astrometric data across multiple lunations during the 2018 observing window. These observations were obtained from SOAR with the Goodman instrument on the three nights of 18, 19 and 20 October 2018 for both SN16 and RY7, and from Lowell Observatory's 4.3-m Discovery Channel Telescope with the Large Monolithic Imager (LMI) on 9 and 12 December 2018 for RY7. The later were obtained when the target magnitude was approximately $V\sim24.5$ and helped to extend its observational arc from 26 to 89 days. These astrometric data were submitted to the Minor Planet Center and are available through their database.

%
%
\section{Physical Characteristics}
\label{sec.properties}

Spectroscopic and photometric observations of the pairs 2015 EE7 -- 2015 FP124 and 2017 SN16 -- 2018 RY7 were used to constrain their physical properties. The spectra are diagnostic of taxonomy and rough composition. The photometry provides loose constraints on the rotational states and shapes of EE7 and FP124. Thermal data of EE7 from the Spitzer Space Telescope were also analyzed to investigate the possibility of a dust environment.

\subsection{Spectral Properties}
\label{sec.spec}

\begin{figure}[t!]
\begin{center}
\includegraphics[width=5in]{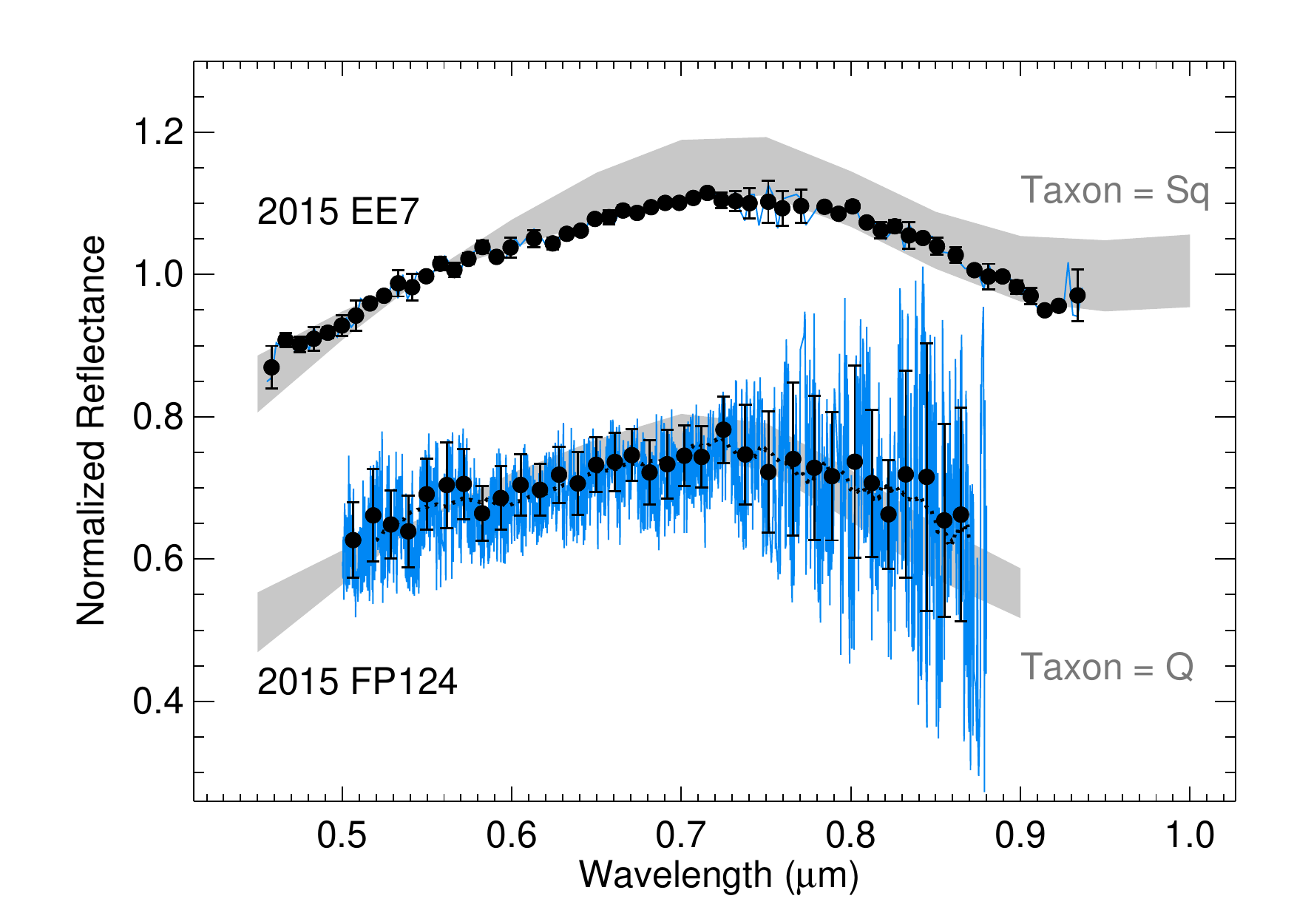}
\caption{Visible wavelength spectra of 2015 EE7 and 2015 FP124 from Gemini-South and SOAR respectively. Data for each object have been offset for clarity. A sigma clipped version of the unbinned data are shown with a dark blue line and the binned data with black dots. A running box car average of the FP124 spectrum is shown as a dashed black line. The envelopes of the best fit taxonomic types for each spectrum are shown in grey in the background. Both objects are in the S-complex, with a preference towards specific types that have experienced low degrees of space weathering.}
\label{fig.spec_ee7}
\end{center}
\end{figure}

Taxonomic classifications for all spectra were achieved by finding the minimum chi-squared residual relative to each of the spectral type envelopes in the \citet{bus02} system. The \citet{bus02} taxonomic envelopes were resampled to the binned resolution of our individual spectra to determine the best fit. Both EE7 and FP124 have spectra indicative of an S-complex classification (Figure \ref{fig.spec_ee7}). While the Gemini spectrum of EE7 is very high quality and clearly indicates an Sq-type classification, the signal-to-noise of the FP124 spectrum is much lower due to poor observing conditions and instrumental fringing at long wavelengths. Nevertheless, the FP124 data are taxonomically diagnostic and are most consistent with a Q-type classification. Comparison of these two spectra suggest that they are statistically indistinguishable within the error bars. This was determined by fitting error weighted polynomials to the data and then comparing the coefficients of the fits. The resulting coefficients were indistinguishable at the 1-sigma level.

The spectral classifications for EE7 and FP124 are consistent with a composition analogous to ordinary chondrite meteorites. It is interesting that these two objects have similar spectral properties with taxonomic types that are indicative of relatively fresh, unweathered surfaces \citep[e.g.][]{vernazza08}. A surface not heavily altered by space weathering indicates a resurfacing event that occurred recently relative to space weathering timescales. There is clear precedent for this in the Main Belt: \citet{polishook14} measured fresh S-complex reflectance spectra for some Main Belt asteroid pairs. Moreover, they noticed that in such cases the smaller secondary displays a fresher spectrum, which may support a model of repeated disruption events for the secondary \citep{jacobson11}. Unfortunately, the timescale for space weathering of ordinary chondrites is not well constrained with estimates ranging from under 1 Myr \citep[e.g.][]{vernazza09} to more than 1 Gyr \citep[e.g.][]{willman10}. If the resurfacing event for this pair is related to their formation, then determination of the age of the system could provide important insight into the space weathering of ordinary chondrites and details of the surface alteration process \citep[e.g.][]{polishook14c}. We discuss dynamical ages based on orbital integrations in \S\ref{sec.dyn}. These integrations show that despite low Minimum Orbital Intersection Distance (MOID) values of 0.07 AU for both EE7 and FP124, tidal perturbations due to close planetary encounters within the past 5 kyr are not a likely cause of these object's fresh surfaces; a hypothesis that has been posed for fresh, ordinary chondrite-like surfaces in the NEO population \citep{nesvorny05,binzel10}.

\begin{figure}[h!]
\begin{center}
\includegraphics[width=5in]{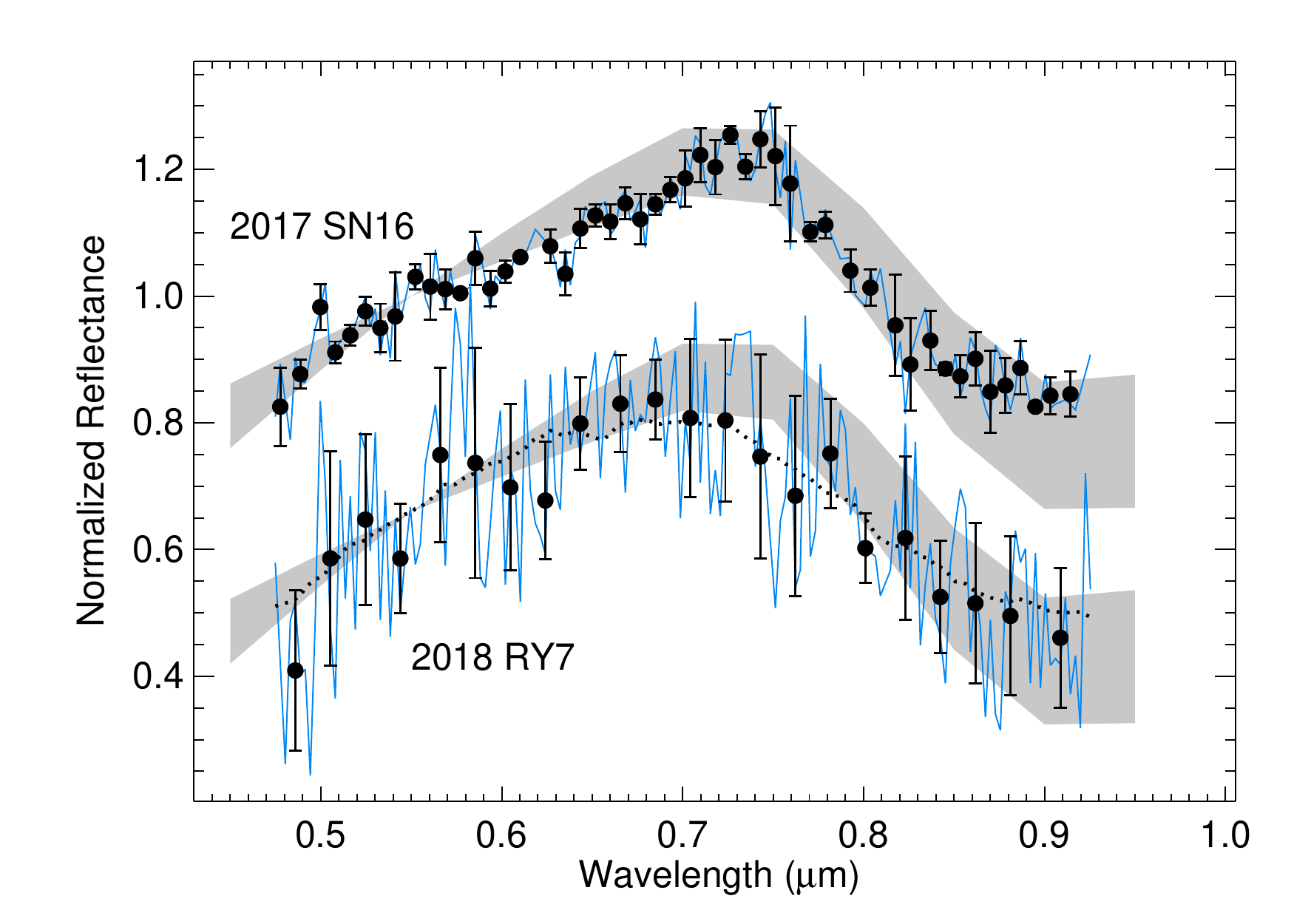}
\caption{Visible wavelength spectra of 2017 SN16 and 2018 RY7 from Gemini-South. Data for each object have been offset for clarity. A sigma clipped version of the unbinned data are plotted with a dark blue line and binned data with black dots. A running box car average of the RY7 spectrum is shown as a dashed black line. The envelopes of the best fit V-type taxonomy are shown in grey in the background.}
\label{fig.spec_sn16}
\end{center}
\end{figure}

Both SN16 and RY7 were best fit with a V-type taxonomic classification (Figure \ref{fig.spec_sn16}). Low signal-to-noise is an issue with both of these spectra, particularly for RY7, but the V-type classification helps to minimize ambiguity in assignment because this type displays the deepest 1 $\mu m$ absorption feature in the \citet{bus02} system and thus is difficult to confuse with other taxonomic types. V-type asteroids are canonicaly associated with basaltic achondrite meteorites and may be collisional fragments of the large Main Belt asteroid 4 Vesta \citep[e.g.][]{moskovitz10}. Unlike asteroids with ordinary chondrite-like compositions, V-types are not expected to display pronounced spectral changes due to space weathering \citep{pieters12}, thus the formation age of this pair is less relevant to understanding space weathering than to constraining dynamical processes in near-Earth space (\S\ref{sec.dyn}). 

The spectra of both pairs are indicative of volatile-poor compositions. This is an important piece of evidence that we use in \S\ref{sec.discussion} to argue for a likely formation mechanism for these systems.

The fact that the individual bodies in these pair systems are each of the same spectral complex is evidence for a common origin, and that the identification of these pairs is unlikely to be a consequence of random dynamical associations. S-complex asteroids represent $\sim$50\% of the observed NEO population \citep{perna18,binzel19,devogele19}, thus EE7 -- FP124 do not offer strong constraints on the probability of two random asteroids in near-Earth space having the same spectral type. \citep[Debiased estimates also find a high fraction $\sim$40\% of NEOs are in the S-complex,][]{stuart04}. However, the V-type classifications for SN16 and RY7 provide stronger constraints on the likelihood of a non-random association. The observed fraction of V-type NEOs is $\sim$2-4\% \citep{perna18,binzel19,devogele19} and the debiased fraction is lower $\sim$1\% \citep{stuart04}. Thus there is only a $\leq0.2\%$ chance of finding two V-type NEOs if randomly selecting any two objects from the population. This suggests at greater than 3-sigma significance (99.8\%) that SN16 and RY7 are a non-random spectral pair. This simple calculation does not take into account the fact that the large majority of V-types likely reach NEO orbits via specific escape routes from the Vesta family in the inner Main Belt, and thus may not necessarily be uniformly distributed in near-Earth space. This would imply that our 3-sigma result could be an upper limit, though further analysis of the taxonomic distribution of NEOs as a function of orbital elements would be required to address this in greater detail.

\subsection{Rotational and Thermal Properties}
\label{sec.rot}

Our attempts to collect lightcurve photometry were less successful, with the data obtained proving to be marginally diagnostic of rotation state. No photometry was obtained for SN16 or RY7. Photometry of FP124 from SOAR on 6 April 2015 (Table \ref{tab.obs}) was of low quality with typical errors of 0.1-0.3 magnitude per data point and was taken in a crowded field around -6$^\circ$ galactic latitude. Despite observing the target for roughly 2.5 hours, only 33 x 7 second exposures of the asteroid were useful for photometry. Heavy extinction over a large fraction of this observing window resulted in non-detection of the target in most frames. No statistically significant periodic signature (based on a Lomb-Scargle analysis) was seen in these data, thus we were unable to place a constraint on the rotation period of FP124. If the underlying rotation period of FP124 is of order a few hours, then we can say that its lightcurve amplitude at this apparition was less than about 0.6 magnitudes based on the scatter in our data.

\begin{figure}[h!]
\begin{center}
\includegraphics[width=6.in]{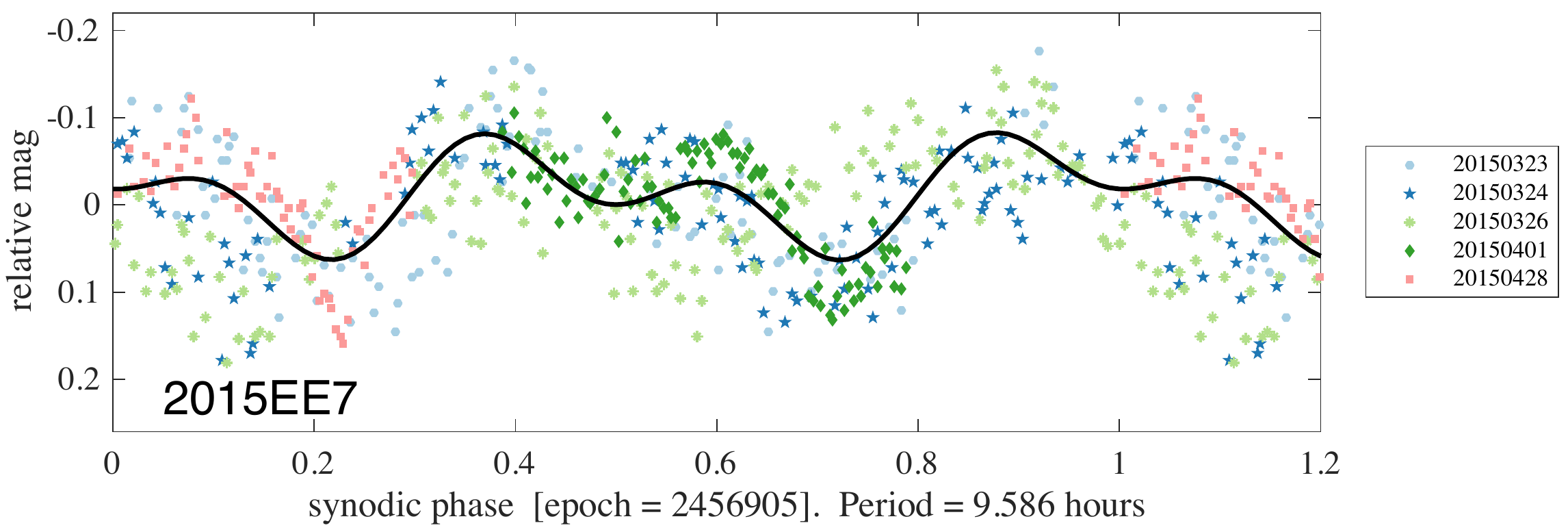}
\caption{Phase folded photometric lightcurve of 2015 EE7. Combination of our data with that from \citet{warner15} suggests a best fit period of 9.586 $\pm$ 0.007 hours and a low amplitude of 0.14 magnitudes. These lightcurve parameters are not strongly constrained, leaving open the possibility of alternative rotation states, including non-principal axis rotation.}
\label{fig.lc_ee7}
\end{center}
\end{figure}

Published photometry of EE7 based on data taken on three nights in late March 2015 suggest a rotation period of 4.71 or 9.42 hours \citep{warner15}. We obtained an additional three nights of lightcurve photometry of EE7 from Lowell Observatory's 1.8-m Perkins telescope and SOAR (Table \ref{tab.obs}). Ultimately we excluded from our analysis the 31 March observations from the Perkins telescope because the scatter in the data were comparable to the amplitude of the lightcurve. Both the \citet{warner15} data and our data were combined for analysis. Our Perkins and SOAR data were resampled to a time resolution of $\sim140$ seconds in an attempt to reduce point-to-point scatter across the individual 15-35 second exposures and to better match the cadence of the \citet{warner15} observations. We employed a Fourier series analysis to fit the lightcurve of EE7 across a range of spin frequencies from 4 to 12 hours per cycle, which was based on apparent intra-night variability. We see no evidence in our data of periods less than $\sim$4 hours. The size of the data sample and the low amplitude of the lightcurve made it difficult to examine periods longer than 12 hours. For a given frequency, $f$, least-squares minimization was used to derive a $\chi^2$ value of the lightcurve fit. The frequency with the minimum $\chi^2$ was identified as the best fit spin rate. The uncertainty was defined by all rates with $\chi^2$ smaller than the minimum $\chi^2$ +  $\Delta\chi^2$, where $\Delta\chi^2$ was calculated from the inverse $\chi^2$ distribution at 3-sigma assuming a frequency for a lightcurve with 2, 3, 4, or 6 harmonics \citep[for more details see][]{polishook14b}. We settled on a 4th order fit (as did \citet{warner15}) as a compromise between tracing the higher order structure in the lightcurve while not overfitting the noise. The rotation period P = $2\pi/f$ with the lowest $\chi^2$ was found to be 9.586 $\pm$ 0.007 hours with a reduced $\chi^2$ of 1.39. The other period considered by Warner (2015), 4.71 hours, gave a higher reduced $\chi^2$ of 2.36, therefore it is less likely, though additional measurements would be useful to completely rule out this shorter period. The amplitude of the best fit lightcurve was 0.14 magnitude (Figure \ref{fig.lc_ee7}). Typical RMS scatter of the photometry relative to the best fit was $\pm$0.1 magnitude.

We also extended the analysis of the thermal infrared photometry of EE7 obtained by the Spitzer NEOLegacy project (\S\ref{sec.ee7}). As part of the Spitzer survey operations the time series of photometry was made publicly available (http://nearearthobjects.nau.edu/neosurvey/lightcurves/2015\_EE7.pdf). These data do not show any clear indication of repeated periodicity, but do indicate a small peak in brightness during the 8000-second imaging sequence, consistent with a low amplitude, multi-hour lightcurve period. As part of our new analysis we created a co-moving stack of all 80 individual images to investigate the possibility of fragments and a dust coma around the body (as might be expected for a recently disrupted object). The 3 sigma point-source detection limit in the co-moving image is 2 $\mu$Jy, which translates into a 3 sigma upper limit for fragment diameters in the co-moving image of $\sim35$ meters. While no obvious coma or fragments were observable around the target, we estimated an upper-limit dust production rate from the co-moving image. Based on the point source detection limit, measured within a circular aperture with radius 2.59", and using the formalism adopted by \citet{mommert14}, we find a 3 sigma upper-limit $(A f \rho) < 10^{-3}$ cm \citep{ahearn84} and a 3 sigma upper limit on potential dust production of $3\times10^{-3}$ g/s. $(A f \rho)$ is estimated from the 3 sigma flux density upper limit assuming a purely thermal nature of that emission. The upper limits account for the observational circumstances, typical dust properties \citep{mommert14}, and a dust particle velocity of 10 m/s, which can be considered an upper limit for dust ejection velocities from active asteroids. The strict upper limits on surrounding fragments and dust production support the absence of any kind of activity in EE7.

It is clear from these data that EE7 displays a low amplitude lightcurve, either because it is a nearly spherical body or because the observations were obtained when the line of sight was aligned closely to the spin axis of the asteroid. With these data we can not distinguish between these possibilities. While we formally obtain a lightcurve period of about 9.6 hours, this period is not strongly constrained by the data. Clearly additional observations are needed. However, it is fair to conclude that any rotation period for EE7 must be longer than $\sim4$ hours. This can be seen in the data from a number of individual dates (e.g. 1 April 2015) where no repeated photometric variability is seen across $\sim4$ hours of continuous observation. In addition, we can not rule out the possibility of a non-principal axis (NPA) rotation state. Certainly the multi-hour variability is consistent with an increase in the occurrence of NPA rotators at slow rotation rates for objects of this size \citep{warner09}, and poor matches in the phasing of lightcurve extrema (e.g. the apparent minimum around phase = 0.7 in Figure \ref{fig.lc_ee7}) are suggestive that a more complicated lightcurve solution is possible. However, these mismatches in phase could simply be due to the changing observing geometry throughout the March - April observing window. Possible large viewing geometry changes can be seen in the evolution of solar phase angle $\alpha$, ecliptic longitude $\lambda$, and ecliptic latitude $\beta$ in Table \ref{tab.obs}. No attempts were made to correct for phase angle effects such as a change in lightcurve amplitude. Again more data are needed to diagnose the possibility of NPA rotation, which would make EE7 the first known asteroid pair primary in a NPA state \citep{pravec19}.

%
%
\section{Dynamical Analysis}
\label{sec.dyn}

The dynamical proximity of the members in these pairs is suggestive of a common origin and, due to the short coherence time of NEO orbits \citep[$\sim10$ kyr, ][]{schunova14}, could be a consequence of a recent break-up event. This possibility merits a more detailed look at the dynamical properties of these systems.

\subsection{Pair Proximity}

We first consider whether the proximity of these objects could be a consequence of random fluctuations in the parameter space of NEO orbital elements. We assess proximity in a quantitative manner using the dimensionless $D$-criterion introduced by \citet{drummond81} to determine associations between the orbits of meteor showers and their comet parent bodies. We adopt this specific metric in lieu of others because it has been shown that the Drummond $D$ criterion provides a robust preservation of orbital proximity for asteroid pairs (albeit in the Main Belt) over time intervals of order 100 kyr \citep{rozek11}. Though there are known issues with the various implementations of $D$ criteria \citep[e.g.][]{jopek93}, those are unlikely to affect the simple analysis presented here. A more detailed analysis of using distance criteria to better identify asteroid pairs in near-Earth space will be the focus of future work.

\begin{figure}[t!]
\begin{center}
\includegraphics[width=5in]{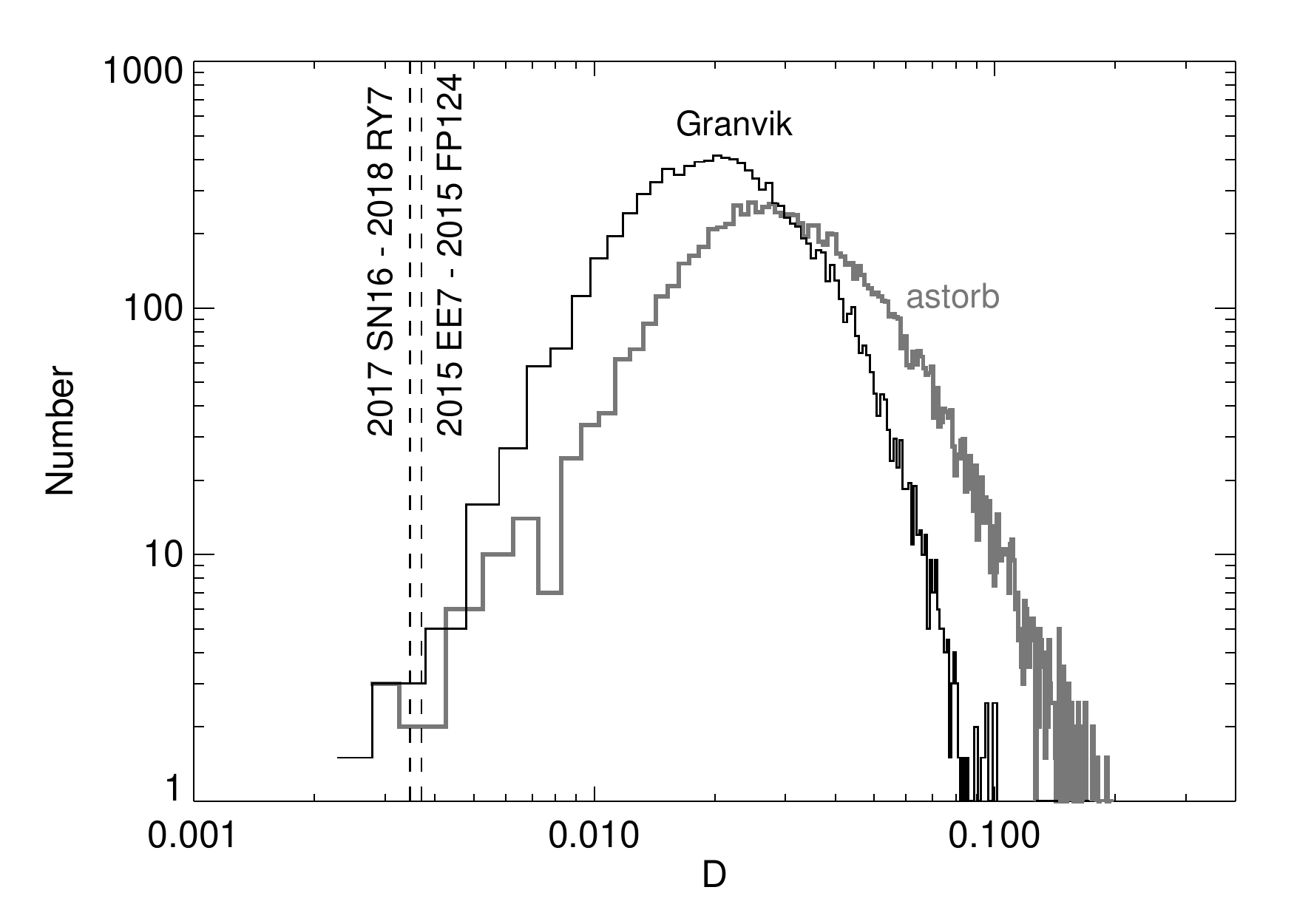}
\caption{Distributions of $D$ criteria for all known NEOs, as represented by Lowell Observatory's {\it astorb} catalog, and 20,000 randomly selected objects from the synthetic NEO population model of \citet{granvik18}. These $D$ values were computed for the nearest neighbor to each object. The values for EE7 -- FP124 and SN16 -- RY7 are $\sim3.5\sigma$ outliers from these distributions, suggesting that such orbital proximity is unlikely to be a consequence of random fluctuations in the orbital element catalog.}
\label{fig.dcrit}
\end{center}
\end{figure}

The Drummond $D$ criteria for EE7 -- FP124 and SN16 -- RY7 are 0.0037 and 0.0035 respectively based on the elements in Table \ref{tab.orb}. To place these in context we compare these values to an ensemble of $D$ values computed from the known NEO population and an ensemble from the \citet{granvik18} NEO population model (Figure \ref{fig.dcrit}). At the time of analysis the known NEO population consisted of 19,615 objects cataloged in Lowell Observatory's {\it astorb} database accessible at https://asteroid.lowell.edu. The Granvik model contains 731,683 synthetic objects that represent the de-biased NEO population up to an absolute magnitude $H=25$. For each of the 19,615 known NEOs we computed $D$ criteria relative to all of the other NEOs in the catalog and then used those to identify a nearest neighbor based on minimum $D$. The same nearest neighbor approach was applied to the full Granvik model for 20,000 randomly selected objects. The distributions of these minimum $D$ criteria are shown in Figure \ref{fig.dcrit}. The Granvik distribution represents the ideal case of an NEO catalog complete down to $H=25$, while the {\it astorb} distribution includes a more realistic assessment that includes discovery bias and incompleteness in the real catalog.

We can estimate the probability that the $D$ criteria for EE7 -- FP124 and SN16 -- RY7 are the outcome of random chance by considering the fraction of $D$ values in the distributions that are less than or equal to those of our objects. In the case of both the synthetic Granvik distribution and the known NEO population we find that there is only a 0.02\% chance that any random pair would have $D$ values as low as our objects. Therefore, our objects are more than $3.5\sigma$ outliers relative to the distributions for the real and synthetic populations, suggesting that their orbital proximity is very unlikely to be the consequence of random fluctuations in the catalog of orbital elements. Re-running this analysis with an $H<25$ threshold applied to the {\it astorb} catalog so that it more closely mimics the Granvik model shifted the {\it astorb} peak to slightly higher $D$, indicating that the smallest known NEOs ($H>25$) skew the distribution towards smaller $D$. This is not surprising given the discovery bias in favor of low geocentric distances for small NEOs. The offset in the peaks between the real and synthetic distributions in Figure \ref{fig.dcrit} is likely a consequence of incompleteness in the catalog of known NEOs.

The probability of 0.02\% for random identification of pairs with $D<0.0035$ in both {\it astorb} and the Granvik catalog is noteworthy. The Granvik $D$ distribution is computed against the full catalog of 731,683 synthetic objects. Thus, to first order it predicts what the real distribution would look like if the known NEO catalog was complete to $H=25$. As such we might expect that incompleteness in {\it astorb} would result in an even lower probability for randomly finding pairs in the known NEO population. The fact that it is not lower may be a consequence of the purely dynamical nature of the Granvik model that does not account for evolutionary processes such as YORP spin-up and fragmentation or thermal disruption.

This analysis highlights an interesting aspect of the NEO catalog related to candidate pairs with very low $D$ criteria. As a demonstration we consider the objects 2005 TE49 and 2017 TV1, which have a low $D$ value in {\it astorb} of 0.0037. However, these objects have orbit condition codes (a metric on the uncertainty associated with a given orbit solution, for details see http://minorplanetcenter.net/iau/info/UValue.html) in the JPL Small-Body Database of 7 and 6 respectively. These high values mean that the orbits of these objects are poorly constrained. In fact, within 1-sigma uncertainties the orbital elements for these two objects are indistinguishable. As such we suggest that these are the same object that have yet to be linked by the Minor Planet Center. It is unclear how many candidate pairs in the NEO catalog with very small D are in fact the same object that have yet to be linked. Fortunately this does not affect our results and instead suggests that the pairs studied here would be even further outliers if this linkage problem were resolved.

\subsection{Orbital Integrations}

A follow-up to addressing the likelihood of non-random association is to assess the age or time of separation of these pair systems. We approached this by performing backwards orbit integrations to search for convergence events sometime in the recent past. Following the methodology of \citet{zizka16} these integrations were performed with 500 clones that randomly sampled each object's orbital errors based on their covariance matrices in the JPL Small-Body Database (https://ssd.jpl.nasa.gov/sbdb.cgi). JPL orbit solutions from 2017 September 15 and 2017 April 6 were used for EE7 and FP125 respectively. These elements corresponded to epoch 2457114.5 for EE7 and 2457114.5 for FP124. All 500 clones of FP124 were first integrated the 6 days to match the starting epoch of EE7, and then the full suite of all 1000 clones were integrated together. Orbit solutions from 2018 December 13 were used for SN16 and RY7. The epoch of elements was 2458283.5 for SN16, and 2458409.5 for RY7. The 500 clones of RY7 were integrated 126 days to match the starting epoch of SN16 and then the full integrations were begun. All clones were integrated backwards using the IAS15 integrator from the python-based REBOUND package \citep{rein15}. IAS15 is a high accuracy non-symplectic integrator that was run with an adaptive timestep initially set to -0.001 years. A convergence event would involve both a small distance between objects and a low relative velocity. We quantified distance by computing the Minimum Orbital Intersection Distance (MOID) between two orbits and relative velocity as the difference of two object's velocity vectors. The relative velocities were computed as if the objects were at the MOID configuration and did not take into account integrated positions (i.e. mean anomaly). MOID and relative velocity were computed for every 15 years of integration. For an asteroid break-up event, the integrated MOID and relative velocity may be expected to trend towards zero at a common time (though a collisional origin would be associated with high ejection velocities). We expect small non-zero values for the velocity and MOID largely due to uncertainty in the integrations caused by orbit errors for the individual objects and chaotic orbit evolution in near-Earth space. In the case of pair formation via YORP spin-up or dissociation of a binary the minimum MOID would likely represent the Hill radius of the primary and the minimum velocity would be similar to the escape velocity of the primary \citep{scheeres09,jacobson11}. For the pairs considered here these values are small: for example the Hill radius of SN16 is $\sim5$ km and its escape velocity is $<0.1$ m/s. We can not resolve such small values given uncertainties inherent to our integrations.

\begin{figure}[h!]
\begin{center}
\includegraphics[width=5.5in]{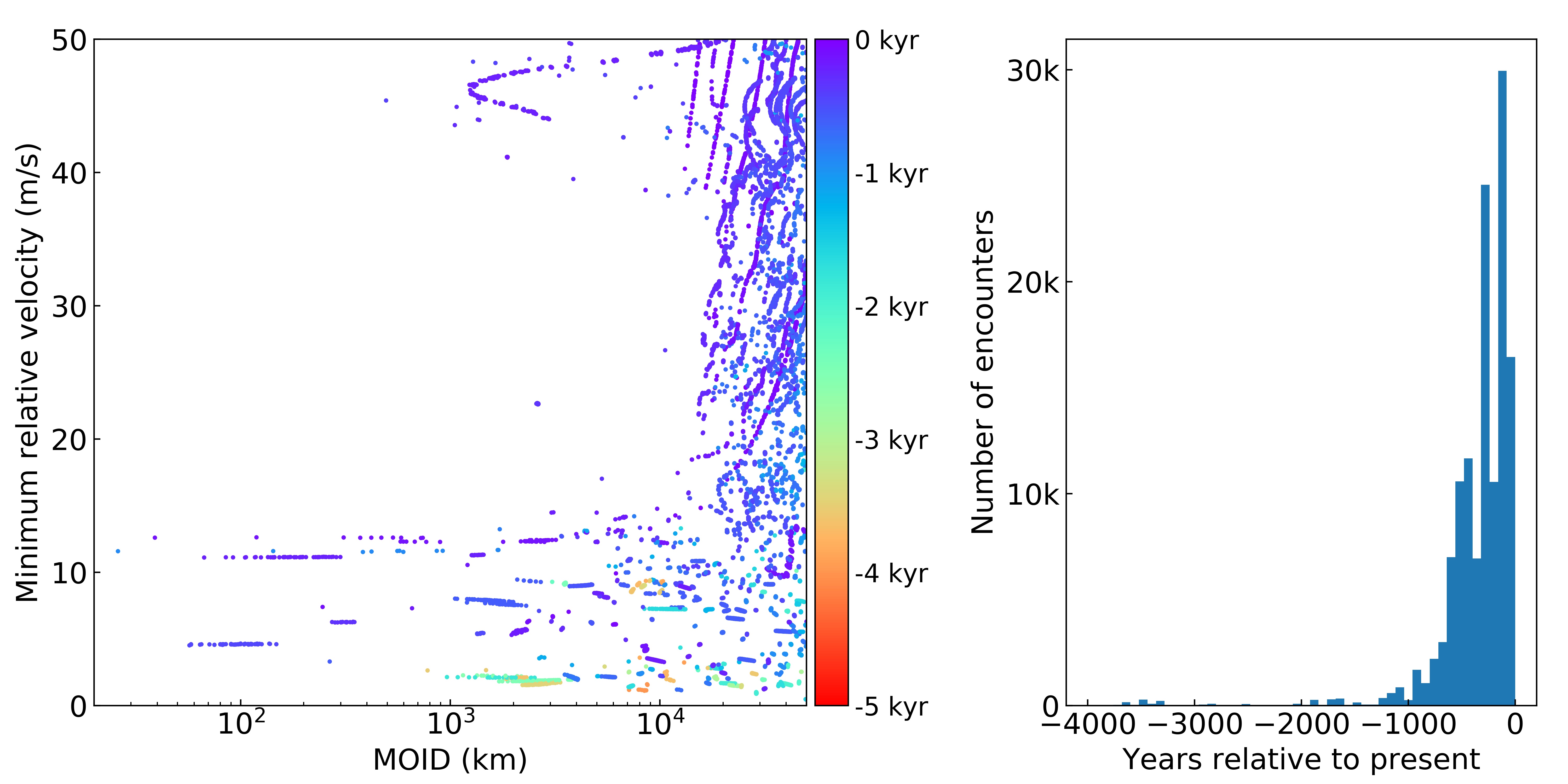}
\caption{Left: The minimum relative encounter velocity within the 5,000 year integration window for 100,000 clone pairings of 2015 EE7 and 2015 FP124. The velocities are plotted as a function of the MOID at the time of each encounter. The color coding of the points indicates the timing of the encounters. Right: Histogram of the times of the minimum velocity encounters. Most encounters occur within the past 1,000 years, however due to large orbital errors the range of possible encounters spans several orders of magnitude in both MOID and velocity.}
\label{fig.ee7_moid}
\end{center}
\end{figure}

Our integrations show that the orbital elements $a$, $e$, $i$, $\Omega$, and $\omega$ (Table \ref{tab.orb}) for all four objects of interest (EE7, FP124, SN16, and RY7) evolve smoothly and are deterministic over the past 5-10 kyr, showing no planetary encounters that would significantly alter their evolution. Unfortunately, the observed orbital arc of FP124 only spans 15 days, thus the errors on its nominal orbital parameters are large. This poor quality orbit translates to large uncertainty in the integrated mean anomaly of FP124 that grows to the full extent of the orbit in $<100$ years. By contrast uncertainty along the orbit for EE7 stays within half of a degree over the past 5000 years. This uncertainty makes it infeasible to reasonably constrain a time of separation for this pair. We do note that throughout the 5 kyr of integrations rare instances of both small MOID ($<100$ km) and low relative velocities ($\sim$few m/s) for orbital clones of this pair can be found at the same time. This is shown in Figure \ref{fig.ee7_moid} where we recorded for 100,000 randomly selected clone pairs the minimum relative velocity that occurred during the 5,000 years of integration, as well as the time and MOID at that minimum velocity configuration. In general, these clone pairs show a wide range of velocity (1-50 m/s) and MOID values ($10^1 - 10^5$ km) at their minimum velocity configurations. The times associated with these minimum velocity configurations are skewed towards the most recent 1,000 years, however this is likely just a consequence of growing orbital uncertainty causing increased scatter in outcomes at more distant times in the past. For example, the scatter in perihelion distance amongst the 500 clones of FP124 grows to nearly 10\% at 5,000 years in the past. As such we limit the integrations for this pair to just the past 5 kyr. Until the orbit quality of FP124 can be improved these integrations are thus not uniquely diagnostic of a specific convergence event.

\begin{figure}[t!]
\begin{center}
\includegraphics[width=5.5in]{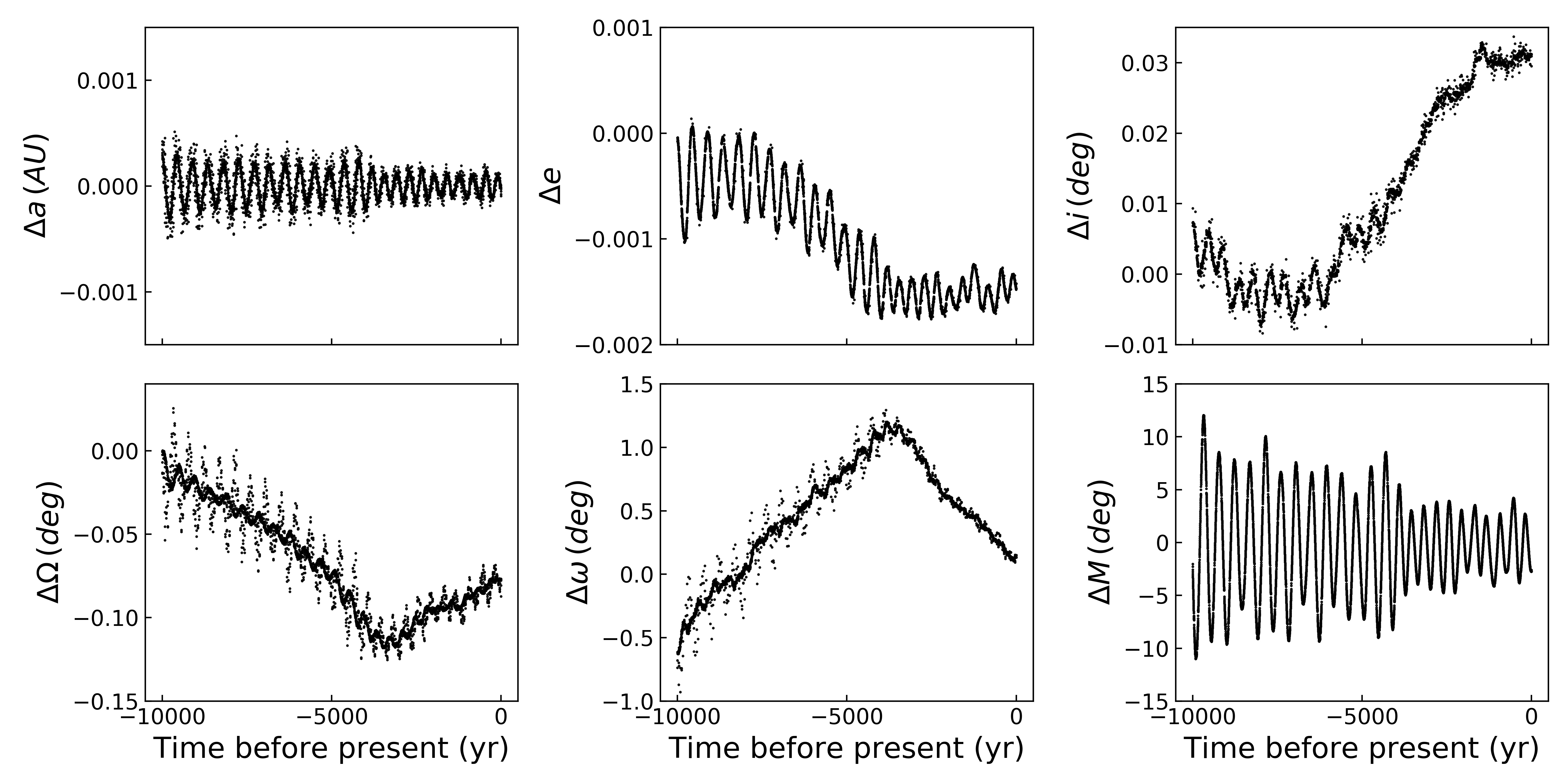}
\caption{Time evolution of orbital element differences (clockwise from upper left: semi-major axis, eccentricity, inclination, mean anomaly, argument of perihelion, longitude of ascending node) for 2017 SN16 and 2018 RY7 based on the nominal orbit solutions for each object. The secular evolution of the orbits is smooth over these integrations and we find that the differences between all orbital elements approach or cross zero some time between 8,000 and 10,000 years ago.}
\label{fig.orbits}
\end{center}
\end{figure}

To mitigate such non-diagnostic results for SN16 and RY7, we made a concerted effort in 2018 to extend their orbital arcs. Targeted observations of both objects from SOAR and DCT extended the arc of SN16 to 391 days and the arc of RY7 to 89 days. In the JPL Small-Body Database this resulted in orbit condition codes of 1 and 4 respectively. In contrast orbit condition codes for EE7 and FP124 are 3 and 8 respectively. These astrometric observations significantly helped in deriving deterministic orbital histories. For SN16 -- RY7 we find that not only are the orbits stable over 10 kyr, but that the two objects track one another closely  (Figure \ref{fig.orbits}). This figure shows the difference in orbital elements based on the nominal orbit solutions (i.e.~not the clones) for these two objects over the past 10 kyr. These differences are generally quite small ($<1\%$) and approach zero for all elements in the range 8,000 to 10,000 years before the present. Roughly the same evolutionary paths are seen for all clones as well. For example, we see  $<0.01^\circ$ scatter in the longitude of the ascending node across all 500 of the clones of 2018 RY7. In addition to the slow secular evolution of the orbits we see a clear $\sim500$ year periodicity in the evolution of each orbital element.

\begin{figure}[h!]
\begin{center}
\includegraphics[width=5.5in]{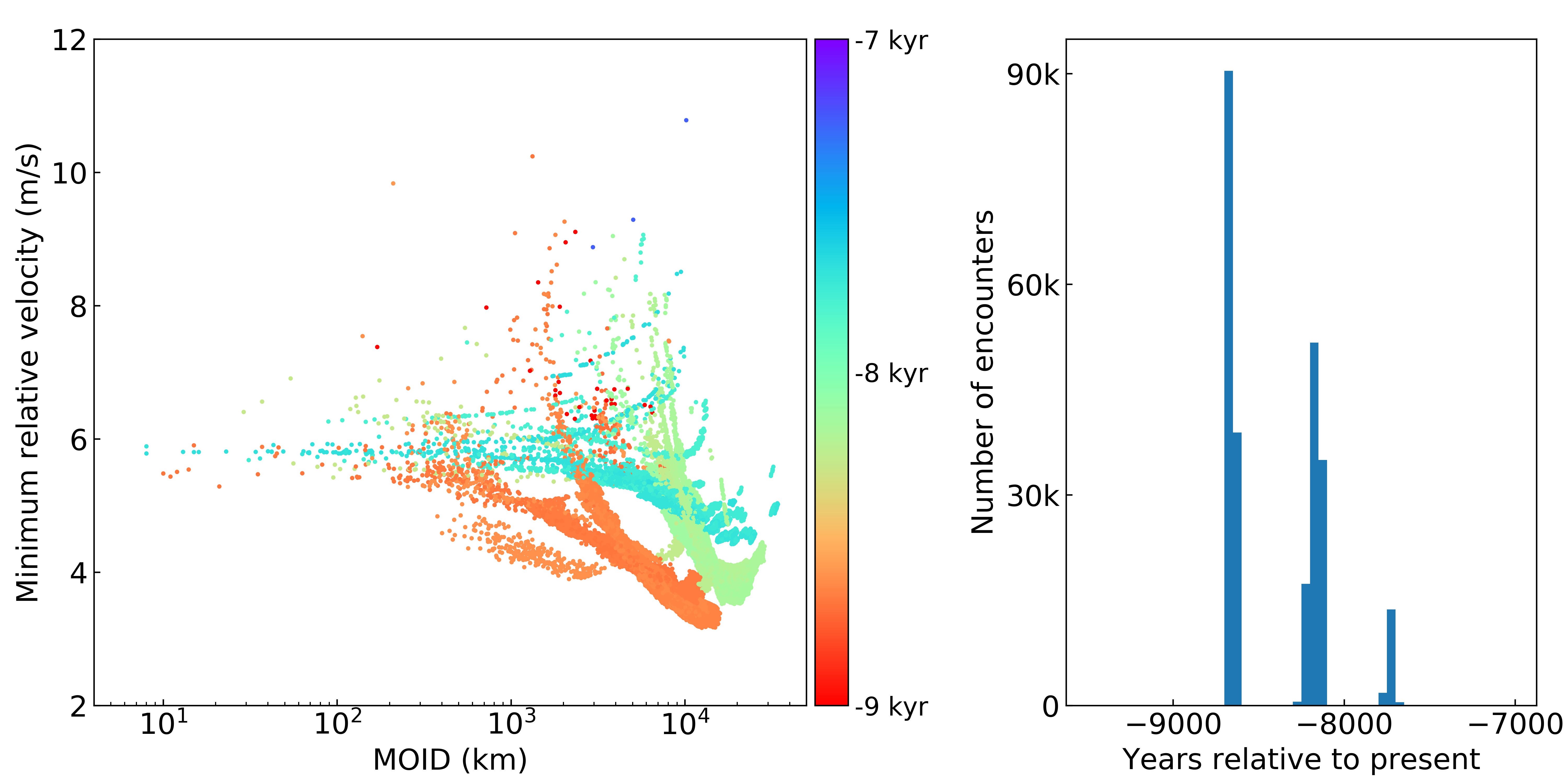}
\caption{Left: The minimum relative encounter velocity within the 10,000 year integration window for all possible clone pairings of 2017 SN16 and 2018 RY7. The velocities are plotted as a function of the MOID at the time of each encounter. The color coding of the points indicates the timing of the encounters. Right: Histogram of the times of the minimum velocity encounters. All encounters occur within a narrow window between 7,500 and 9,000 years before the present.}
\label{fig.sn16_dyn}
\end{center}
\end{figure}

To further probe the possibility of a convergence event within the past 10,000 years, we analyzed all of the 250,000 possible pairings of the 500 clones for SN16 and for RY7 (Figure \ref{fig.sn16_dyn}). As before we recorded the minimum relative velocities between clones that occurred during the full integration, as well as the time and MOID at that minimum velocity configuration. All clone pairings show small minimum velocities of $\sim3-10$ m/s and MOID values at the minimum velocity configuration generally $\sim10^2-10^4$ km. A subset of these clone pairs show MOID values less than 10 km at the time of their minimum velocity encounters. There are frequent times in these integrations when smaller MOID values are seen, even down to values $<1$ km, however those smaller MOIDs typically occur when relative velocities are higher $\sim10-40$ m/s. We find that the timing of all minimum velocity encounters are clustered within a narrow window of about -7,500 to -9,000 years before present (Figure \ref{fig.sn16_dyn}, right panel). These results suggest that SN16 -- RY7 may be an extremely young pair, having separated within the last 10 kyr.

\begin{figure}[h!]
\begin{center}
\includegraphics[width=3.25in]{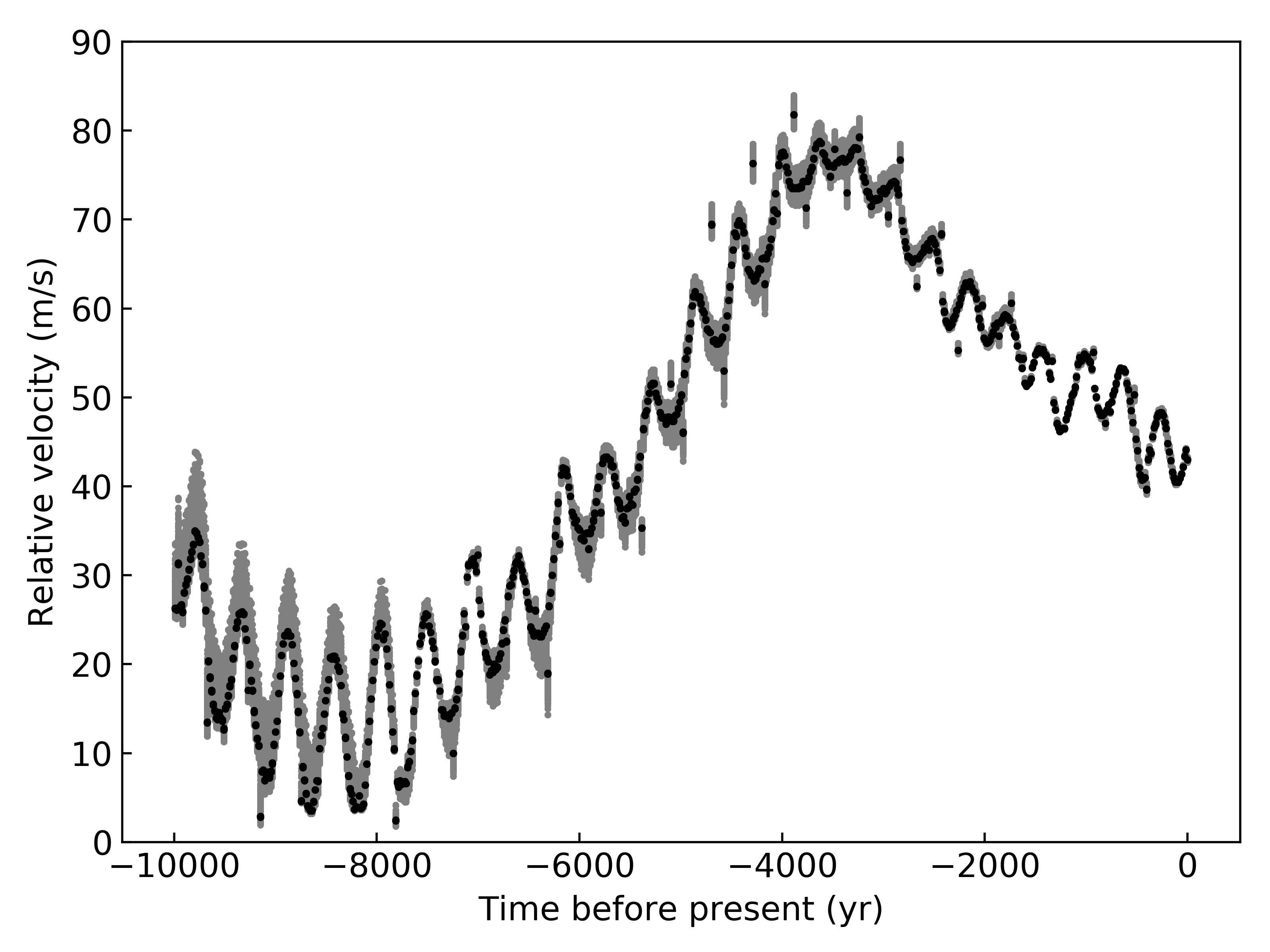}
\includegraphics[width=3.25in]{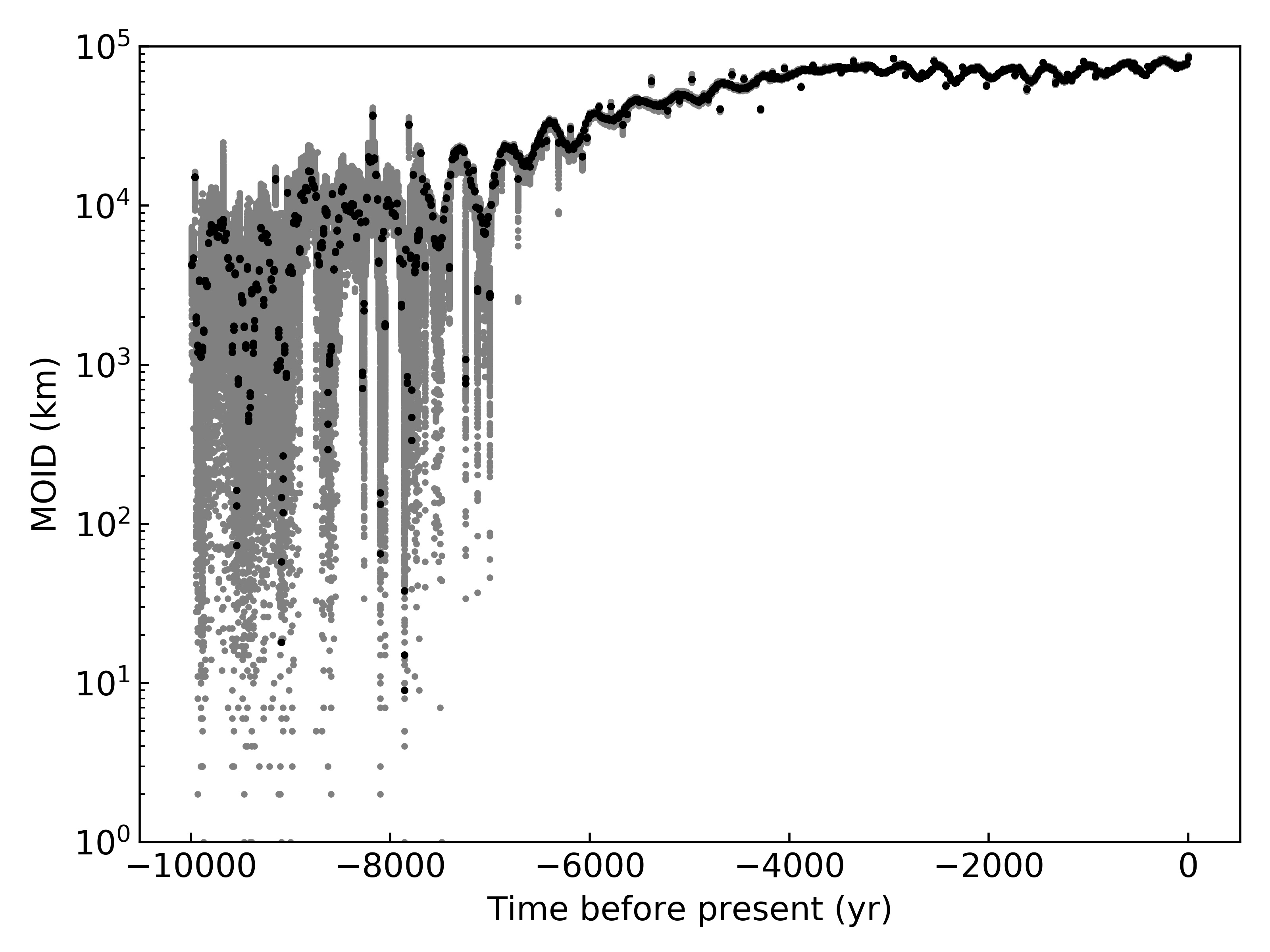}
\caption{Relative velocity (top) and MOID (bottom) evolution for the nominal orbit of SN16 -- RY7 (black) and 500 randomly selected clone pairs (grey). Velocities and MOID have been computed every 15 years throughout the integration. The 500 year periodicity, also seen in the orbital element evolution, is apparent here. The three velocity minima at -7700, -8200 and -8700 years are roughly the same depth and are the cause of the discrete timing of minimum velocity encounters in Figure \ref{fig.sn16_dyn}. Values of very low MOID ($<10$ km) appear at times before -7,000 years.}
\label{fig.sn16_vrel}
\end{center}
\end{figure}

The histogram of clone encounter times shows three distinct spikes centered around 8,500 years ago. These are approximately evenly spaced with about 500 years between them. The evolution of relative velocities for the nominal orbits as well as all clone pairs follow the same 500 year periodicity as the orbital elements (Figure \ref{fig.orbits}). This creates local minima in the relative velocity evolution that are superimposed on an overall secular trend that reaches a single global minimum around -8,500 years (Figure \ref{fig.sn16_vrel}). Three of the local minima (around -7700, -8200 and -8700 years) overlap the minimum of this global trend and all three reach roughly the same minimum velocity values. We interpret the three spikes in the histogram of encounter times (Figure \ref{fig.sn16_dyn}) as a consequence of these local minima in the relative velocity evolution. Very low values of MOID, in some cases reaching values $<10$ km, are apparent in the integrations at times earlier than -7,000 years, however a wide range of values are still possible.

Superimposed on the secular and 500 year variability trends in Figure \ref{fig.sn16_vrel} are discrete spikes that appear as outliers in single timesteps. Two of these spikes towards lower velocity occur around the global minimum and result in relative velocities $<1$ m/s. We do not see an association with low MOID values at these specific time steps, and thus do not interpret these low velocity spikes as evidence for specific convergence events. Instead this analysis supports a convergence event across a range of dates, potentially in the window of -9,000 to -7,500 years ago. Further analysis beyond the scope of this work is needed to determine whether an actual encounter of SN16 -- RY7 can be assigned to just a single set of dates.

Our dynamical analysis does not account for the Yarkovsky effect \citep{vok15}. Even though this is a small perturbation, it becomes relevant over the long timescales considered in this paper. Even so, the analysis presented here remains valid. First of all, the Yarkovsky effect primarily manifests as a runoff in longitude that accumulates quadratically with time. While the location along the orbit will be affected, the MOID and relative velocity calculations do not depend on the mean anomaly. Moreover, the uncertainty in the Yarkovsky related semi-major axis drift increases the uncertainty in the past values of semi-major axis, which allows even smaller values of MOID and relative velocity, as well as the possibility for more recent separation events. In future work we will attempt to identify the timing of a specific separation event for SN16 and RY7, which will necessitate improved knowledge of these object's orbits and and constraints on the Yarkovsky effect acting on these objects.

%
%
\section{Discussion}
\label{sec.discussion}

We have presented the physical characterization and dynamical analysis of two newly identified NEO pairs: 2015 EE7 -- 2015 FP124 and 2017 SN16 -- 2018 RY7. We find that the members within each system are of the same spectral complex. Attempts to determine rotation state with lightcurve photometry were unsuccessful. Based on Spitzer Space Telescope images, no dust emission was detected around EE7 down to an upper limit on the dust production rate of $3\times10^{-3}$ g/s. For one of the pairs, SN16 -- RY7, we were able to extend orbital arcs with new astrometric observations, which improved our knowledge of these object's orbits and thus improved the outcome of our dynamical analysis. This analysis suggests that this pair may be as young as $\sim8000$ years old, making it amongst the youngest known asteroid pairs in the Solar System. Additional spectroscopic, photometric, and astrometric data would be of great value in further constraining the compositional, rotational, and orbital parameters of these objects. 2015 EE7 will next be observable in Fall 2024 at a peak brightness of V=21.8. New lightcurve photometry at this time would help to constrain the rotation period of this object and potentially diagnose whether it is in a non-principal axis rotation state. Due to the short arc for 2015 FP124, this object is effectively lost such that predicting the specifics of future windows of observability (e.g. magnitude and specific location on sky) is not possible. This object will have to be re-discovered in the future. The Large Synoptic Survey Telescope \citep[LSST,][]{ivezic08} may detect FP124 in the Fall of 2024 when the object reaches $V\sim23.5$ far from opposition. Both 2017 SN16 and 2018 RY7 are observable in the Fall of 2019 with peak brightnesses of V=21.5 and 22.2 respectively. Lightcurve photometry at this time would be valuable to constrain these object's rotation periods, which could be diagnostic of the mechanism(s) responsible for their formation \citep{jacobson11}. Additional astrometry at this time will further reduce uncertainties on these object's orbits. The next apparition after 2019 for these objects will not be until the late 2050's.

An interesting aspect of asteroid pairs in near-Earth space is the possibility for formation mechanisms that are not relevant to pairs in the Main Belt. Possibilities for NEO pair formation include collisional disruption of a larger parent body, separation of a binary system, YORP spin-up and rotational fission, tidal disruption from planetary encounters, and/or thermally driven fragmentation possibly facilitated by volatile loss \citep{richardson98,vok08,jacobson11,granvik16}. Our results can exclude several of these possibilities for the pairs considered here. The spectral data suggest taxonomic types (S- and V-type) which are interpreted to be associated with largely dehydrated, volatile-poor compositions. Thus thermal fragmentation driven by volatile loss is not a likely explanation for these two systems.  Our numerical integrations show that none of these objects experienced close planetary encounters in the past 5-10 kyr, thus tidal disruption is not a likely formation mechanism. Furthermore, large scale collisions (i.e. those large enough to fragment $\sim100$-m scale bodies) are infrequent in near-Earth space and not thought to be an important part in the evolution of the NEO population. Therefore, we suggest that like the majority of their Main Belt counterparts \citep{pravec10}, these two NEO pairs formed via YORP spin-up and/or dissociation of binary systems. If YORP played a role in their formation, then we might expect the primary members of these pairs to have the characteristic top-shaped morphology that is thought to be a consequence of radiative spin-up and fragmentation \citep[e.g.][]{walsh08}. The low amplitude lightcurve for 2015 EE7 is consistent with a nearly spherical morphology. With new lightcurve photometry a similarly low amplitude lightcurve might also be expected for 2017 SN16.

The spin state of asteroid pairs can provide important clues about the mechanism(s) of their formation \citep{pravec10,jacobson11,pravec18}. We have presented here loose constraints on the spin state of EE7. The suggested period around 9.5 hours would place EE7 on the upper edge of predicted periods for pairs that formed via YORP spin-up and rotational fission \citep{pravec19}. Our data can not rule out the possibility of a non-principal axis spin state for EE7, and in fact our inability to find a robust single period solution could be an indicator of NPA rotation. There are interesting implications for EE7 if we speculate that it is in such a rotation state. Based on our dynamical integrations we know that EE7 has not experienced recent planetary encounters, thus NPA rotation could be a result of its formation. \citet{jacobson11} suggest that when multiple asteroid systems form via YORP spin-up and fission, the components will likely be in a NPA state. Over time these NPA rotators will damp energy through internal stresses on a timescale that is dependent on parameters such as the internal rigidity and energy dissipation efficiency \citep{burns73}. These parameters are not well constrained. Therefore, detection of a young asteroid pair still in NPA rotation could place meaningful constraints on these fundamental properties. Pairs in NEO space, due to the short coherence time of their orbits, are an ideal laboratory for testing these scenarios of spin state evolution and asteroid dynamics.

\section{Summary}
\label{sec.summary}

The following points summarize the findings presented here and identify future areas of work relevant to asteroid pairs in near-Earth space:

\begin{itemize}

\item Physical characterization data and dynamical integrations suggest that the NEO pairs presented here are in fact genetically related and are not the consequence of random fluctuations in the parameter space of NEO orbital elements. A systematic survey of other candidate NEO pairs would be beneficial to further test genetic relationships. In addition, formalized metrics for identifying new NEO pairs would be valuable to provide a census of this sub-population of NEOs.

\item Spectroscopic data show that the members of the pair systems presented here are of the same taxonomic complex. Both systems have taxonomic types consistent with volatile-poor compositions, which in conjunction with dynamical arguments, suggest that these objects formed via YORP spin-up and/or dissociation of binary systems. Other formation mechanisms like tidal disruption during planetary encounters or thermal disruption aided by volatile loss are less likely.

\item We have presented a plausible separation age of $<10$ kyr for the pair 2017 SN16 -- 2018 RY7. If true these objects would be one of the youngest known asteroid pairs in the Solar System. Based on dynamical analysis of this pair, it is clear that high quality orbits are a necessity for constraining age. In this case orbit conditions codes $\leq4$ were required.

\item Determining the age of pair systems is incredibly valuable for testing a variety of models including those related to space weathering and spin state evolution. The likely young ages of these systems provide a unique opportunity to probe timescales that are not currently represented in other small body populations.

\item A more detailed analysis of orbital integrations with a large number of clones might enable the determination of specific formation ages (as opposed to a range of ages). Extending the astrometric data arc would improve the orbital knowledge necessary to identify possible separation events. Because of the long timescales, identifying separation events would require modeling of the Yarkovsky effect. Though no Yarkovsky estimate is currently available, constraints can be obtained by astrometric and physical observations. 

\item As the NEO catalog continues to grow with current and future discovery surveys, the known population of NEO pairs will also increase. Systematic identification and monitoring of these objects at the time of their discovery will be important to obtain physical characterization data (composition, rotation state, orbit parameters) when these objects are most readily accessible to telescopic study.

\end{itemize}

The identification and characterization of asteroid pairs amongst NEOs is a relatively recent endeavor  \citep[e.g.][]{ohtsuka06,jewitt06,ohtsuka07,kinoshita07} and has been enabled by the rapid growth of the catalog of known NEOs. The continued study of these objects as an ensemble will ultimately provide new insights into the physical properties and evolutionary processes that have relevance to all asteroids across the Solar System.

{\bf Acknowledgements}\\
We are grateful to two anonymous referees for their careful reading of this manuscript, which led to significant improvements. Primary funding support for this work was provided by NASA grant numbers NNX14AN82G and NNX17AH06G (PI N. Moskovitz) issued through the Near-Earth Object Observations program to the Mission Accessible Near-Earth Object Survey (MANOS). We are grateful to Matthew Knight (U. Maryland) and David Schleicher (Lowell Obs.) for using Lowell's Discovery Channel Telescope to obtain images of 2018 RY7 in December 2018. We acknowledge the great value and utility of the Asteroid Lightcurve Database (ALCDEF, alcdef.org) from which we retrieved archival photometry of 2015 EE7. P. Fatka was supported by the Charles University, project GA UK No.842218. D. Farnocchia conducted this research at the Jet Propulsion Laboratory, California Institute of Technology, under a contract with NASA. FED acknowledges funding from NASA under Grant No. 80NSSC18K0849 issued through the Planetary Astronomy Program. J. Snow is grateful to D.S. for assistance beyond the wall. Results presented here were based on observations obtained at the Southern Astrophysical Research (SOAR) telescope, which is a joint project of the Minist\'{e}rio da Ci\^{e}ncia, Tecnologia, Inova\c{c}\~{o}es e Comunica\c{c}\~{o}es (MCTIC) do Brasil, the U.S. National Optical Astronomy Observatory (NOAO), the University of North Carolina at Chapel Hill (UNC), and Michigan State University (MSU), and on observations obtained at the Gemini-South Observatory, which is operated by the Association of Universities for Research in Astronomy, Inc., under a cooperative agreement with the NSF on behalf of the Gemini partnership: the National Science Foundation (United States), National Research Council (Canada), CONICYT (Chile), Ministerio de Ciencia, Tecnolog\'{i}a e Innovaci\'{o}n Productiva (Argentina), Minist\'{e}rio da Ci\^{e}ncia, Tecnologia e Inova\c{c}\~{a}o (Brazil), and Korea Astronomy and Space Science Institute (Republic of Korea).





\end{document}